\documentclass[a4paper,fleqn,usenatbib]{mnras}

\usepackage{newtxtext,newtxmath}

\usepackage[T1]{fontenc}
\usepackage{ae,aecompl}

\usepackage{graphicx,psfig}	
\usepackage{amsmath}	
\usepackage{graphics}
\usepackage{multicol}

\newcommand{\ltsima}{$\; \buildrel < \over \sim \;$}
\newcommand{\simlt}{\lower.5ex\hbox{\ltsima}} 
\newcommand{\gtsima}{$\; \buildrel > \over \sim \;$}
\newcommand{\simgt}{\lower.5ex\hbox{\gtsima}} 

\newcommand{\feka}{\mbox{Fe  K$\alpha$}}
\newcommand{\fekb}{\mbox{Fe  K$\beta$}}

\newcommand{\xmm}{{\emph{XMM-Newton}}}

\newcommand{\lum}{erg~s$^{-1}$}
\newcommand{\flux}{{erg~cm$^{-2}$~s$^{-1}$}}
\newcommand{\nh}{cm$^{-2}$}
\newcommand{\nhsym}{N_{\mbox{\scriptsize H}}}

\newcommand{\chandra}{{\emph{Chandra}}}

\newcommand{\errUD}[2]{\ensuremath{^{+#1}_{-#2}}}

\newcommand{\suzaku}{{\emph{Suzaku}}}

\newcommand{\XMM}{{\emph{XMM}}}

\newcommand{\swift}{{\emph{Swift}}}
\newcommand{\logxi}{erg cm s$^{-1}$}
\newcommand{\nustar}{{\emph{NuSTAR}}}

\newcommand{\sorg}{MCG-03-58-007}

\title{The stratified disk wind of \sorg}

\author[V. Braito et al. ]{V.  Braito$^{1,2}$\thanks{E-mail:valentina.braito@brera.inaf.it},  J . N. Reeves$^{1,2}$, P. Severgnini$^{1}$,  R. Della Ceca$^{1}$,
\newauthor
   L. Ballo$^{3}$, C. Cicone$^{4}$, G.~A. Matzeu$^{1,3}$, R. Serafinelli $^{1}$, M. Sirressi$^{5}$ \\
$^{1}$INAF - Osservatorio Astronomico di Brera, Via Bianchi 46 I-23807 Merate (LC), Italy \&  Via Brera 28, 20121, Milano\\
$^2$Department of Physics, Institute for Astrophysics and Computational Sciences, The Catholic University of America, Washington, DC 20064, USA\\
$^{3}$European Space Astronomy Centre (ESA/ESAC), E-28691 Villanueva de la Canada, Madrid, Spain\\
 $^4$ Institute of Theoretical Astrophysics, University of Oslo, P.O. Box 1029 Blindern, 0315 Oslo, Norway\\
$^5$Department of Astronomy, AlbaNova University Center, Stockholm University, SE-10691 Stockholm, Sweden
 }
\date{Accepted XXX. Received YYY; in original form ZZZ}

\pubyear{2017}

\begin{document}
\label{firstpage}
\pagerange{\pageref{firstpage}--\pageref{lastpage}}
\maketitle

\begin{abstract}

Past \suzaku, \xmm\  and \nustar\ observations of the nearby ($z=0.03233$)  bright Seyfert 2 galaxy \sorg\ revealed the presence of two deep and  blue-shifted iron K-shell absorption line profiles. These could be explained with the presence of two phases of a highly ionized, high column density  
accretion disk wind  outflowing with  $v_{\rm out1}\sim -0.1c$ and $v_{\rm out2}\sim -0.2c$. Here  we present  two new observations of \sorg: one was  carried out  in 2016 with \chandra\  and one in 2018 with \swift. Both  caught \sorg\  in a brighter state ($F_{{\mathrm 2-10\, keV}} \sim 4 \times 10^{-12}$ \flux)   confirming the presence of the fast disk wind. 
The multi-epoch   observations  of \sorg\ covering the period from 2010 to 2018 were then analysed. These data  show that the lower velocity component  outflowing with $v_{\rm out1}\sim -0.072\pm 0.002c$ is persistent and detected in all the observations, although it 
is variable in column density in the range $\nhsym \sim 3-8 \times 10^{23} $\,\nh.  In the 2016 \swift\ observation  we  detected again  the second faster component outflowing with  $v_{\rm out2}\sim -0.2c$, with a column density ($\nhsym =7.0\errUD{5.6}{4.1}\times 10^{23}$\,\nh), similar to that seen during the \suzaku\ observation. However during the \chandra\ observation two years earlier, this zone was not present ($\nhsym <1.5 \times 10^{23}$\,\nh), suggesting that this faster zone is intermittent. 
Overall the multi-epochs observations  show that  the disk wind in \sorg\ is not only  powerful, but also extremely variable, hence placing \sorg\ among unique disk winds such as the one seen in  the famous QSO PDS456. One of the main results of this investigation is the consideration that these winds could be extremely variable, sometime appearing and sometime disappearing; thus to reach solid and firm conclusions about their energetics multiple observations are mandatory.

\end{abstract}



\section{Introduction}

Less than  twenty years after the  discovery  of the first examples of highly-ionized  (log ($\xi / \rm{erg\, cm\, s^{-1})}=3-6$), massive and fast ($v>0.1 c$) outflowing absorbers (PDS\,456, \citealt{Reeves2003}, PG1211+143,   \citealt{Pounds2003} and APM\,08279+5255, \citealt{Chartas2002}),  it is now largely accepted that fast disk winds are commonly observed  in nearby bright  AGN (\citealt{Tombesi2010,Tombesi2012,Gofford2013,Gofford2015}). Although a  lot of  observational and theoretical effort has been invested in understanding their nature  and their possible role in shaping the host galaxies (\citealt{King2010,King2011,Nardini_Zubovas18} and references therein),   several questions still remain open including their main driving mechanism, their variability  and  their real impact on the host galaxy.
The  velocities measured for these winds  can reach up to $\sim 0.3\, c $ (PDS\,456 \citealt{Reeves2009,Reeves2014}, APM\,08279+5255 \citealt{Chartas2002}),  suggesting that they most likely originate in the innermost region of the accretion disk  (see \citealt{King_Pounds2015}) and therefore could be  linked  to  the accretion process itself  (\citealt{King2003,King2010,King_Pounds2015}). They could be driven either by the radiation pressure    (\citealt{Proga2000,Proga2004,Sim2008,Sim2010}), by  magneto-rotational forces (MHD models: \citealt{Kato2004,Kazanas2012,Fukumura2010,Fukumura2017}) or a combination of both. \\
The measured relativistic velocities and high column densities imply that, although affected by high uncertainties,  the outflow rates and kinetic output  can be huge and match or exceed the conventional threshold of  $L_{\rm {KIN}}/L_\mathrm{bol} \sim0.5-5$\% for  an efficient AGN  feedback on the host galaxy (\citealt{HopkinsElvis2010,DiMatteo2005}).  These disk winds might indeed drive the  
 massive molecular outflows     seen on kpc-scales (\citealt{Cicone2014,Cicone2015,Fiore17})   and thus  influence   the  host galaxies by sweeping away   the interstellar medium   and   suppressing  star formation (\citealt{King_Pounds2015,Zubovas_King12,Zubovas_King16}). Thus   these disk winds could play a major role in the feedback process that shapes the formation of the stellar bulges and simultaneously self regulate the growth of the super massive black hole (SMBH),   leading to the observed  AGN-host galaxy relationships like the $M-\sigma$ relation (\citealt{Magorrian1998,Ferrarese2000,Gebhardt2000}).   \\
The first detections of  powerful X-ray  disk winds  in two Ultra Luminous Infrared Galaxies (ULIRGs),  where  massive large-scale molecular outflows are also present (Mrk 231; \citealt{Feruglio2015} and IRASF\,11119+3257; \citealt{Tombesi2015}) seemed to support   a scenario where  the molecular outflows are driven by an energy conserving disk wind. Here,   when  the disk wind   propagates and shocks with the ISM it does not cool efficiently and the  large-scale outflow  receives a momentum boost (\citealt{King2010,Faucher2012}).   However, recent results   on  other  ultra fast  disk winds observed with the Atacama Large Millimeter/submillimeter Array (ALMA)  show  that not all  outflows lie on the energy conserving  relation.  Actually,  with the more recent IRAM and ALMA data,  only Mrk\,231 and IRAS\,17020+4544 (\citealt{Longinotti2018})  are consistent with the above scenario. An energy conserving wind can be clearly ruled out for both  the prototype of the fast disk winds PDS\,456 (\citealt{Bischetti2019}) and  the powerful wind of I\,Zw\,1 (\citealt{ReevesBraito2019}). This hints for a range of efficiencies in transferring the kinetic energy of the inner wind out to the large-scale molecular component (\citealt{Mizumoto2019}), which suggest that the  role of the powerful disk winds  in the    galaxy evolution may be more complex that what we thought.   Several scenarios can indeed explain the lack of a powerful molecular outflow; among them the large scale gas  could  be clumpy as seen in the ALMA observation of PDS456 (\citealt{Bischetti2019}) or it could be   ionized.  Furthermore, once we consider that the typical dynamical timescales of the large scale outflows are of the order of $t\sim 10^6-10^7$\, yr, it is not unexpected that we do not observe the two phases simultaneously. Indeed, there could be a substantial delay between the onset of the fast X-ray wind and the large-scale energy conserving wind (\citealt{King2011}). \\
 \begin{table*}
\caption{Summary of  the observations used: Observatory, Observation Date, Instrument,  Elapsed  and Net exposure times. The  net exposure times  are obtained after the screening of the cleaned event files. 
\label{table:log_observ}
}

\begin{tabular}{cclcc}
 \hline
 Mission &  Start Date (UT Time) & Instrument  & Elapsed Time (ks) &Exposure$_{\rm(net)}$ (ks)\\
 \hline

\suzaku\ & 	2010-06-03 16:50    &XIS  & 187.4  &  87.0 \\

   \xmm\ &  2015-12-06 12:15  &   EPIC-pn&131.3 &  59.9\\
   \nustar\  &2015-12-06 10:36 &FPMA &281.8 & 131.4\\
    
  \chandra\ &2016-09-25 05:14 &ACIS-S &44 & 39.8\\
  \chandra\ &2016-09-20 09:23 &ACIS-S &17 & 15.5\\
\swift\ & 2018-04-15 13:22 &XRT &-& 19.6 \\
\swift\ & 2018-04-19 00:27 &XRT &-& 17.6 \\
\swift\ & 2018-04-20 00:29 &XRT &-& 6.1 \\
\swift\ & 2018-04-22 03:28 &XRT &-& 19.7 \\
\swift\ & 2018-04-23 23:58 &XRT &-& 3.0 \\
\swift\ & 2018-04-25 03:01 &XRT &-& 3.8 \\
 \hline
\end{tabular}
\end{table*}
It is clear that, if we want to establish the role of the AGN disk winds at driving the large scale outflows, we need more examples of disk winds in conjunction with multiwavelength observations of the host galaxies, which can reveal the large-scale component of the outflows, as well as further investigate   the most compelling cases of  X-ray disk winds.    First of all it is essential to  establish if the fast wind is persistent or not,  then  we need to refine the measurement of the persistent component of the wind, so that we do not introduce an error on the estimates of the mass and energetics of the  disk wind. We note, that with the exception of PDS\,456, the variability of the X-ray disk winds  is poorly studied. In the best studied examples the fast disk wind can be variable in ionisation, column density ($\nhsym$) and even velocity  (e.g. PDS~456, \citealt{Reeves2018,Matzeu2017}; IRASF\,11119+3257, \citealt{Tombesi2017}; PG\,1211+143, \citealt{pg1211} and  APM\,08279+5255, \citealt{Saez2011}).  This observational evidence is  in agreement with  disk wind simulations where the stream is not expected to be a homogeneous and constant flow  (\citealt{Proga2004,Giustini2012}).  The observed variations could be either explained with our line of sight intercepting different clumps or streams of the winds or with a response of the wind to the luminosity of the X-ray source.  Recently, a  direct correlation between the outflow  velocity   and the intrinsic ionising  luminosity was  reported for   PDS\,456 (\citealt{Matzeu2017}),  IRAS\,13224-3809 (\citealt{Chartas2018})  and APM\,08279+5255 (\citealt{Saez2011}), while a correlation between the ionisation of the disk wind and the X-ray luminosity was reported for IRAS\,13224-3809  by \citet{Pinto2018}. This indicates the importance of the incident radiation upon the wind: as the luminosity  and thus the radiation pressure increases, a faster wind is driven.  

\sorg\ is a bright and nearby  Seyfert 2 galaxy  ($F_\mathrm{2-10\, keV}\sim 2 \times10^{-12}$\,erg\,cm$^{-2}$\,s$^{-1}$, $z=0.03233$; \citealt{Sirressi2019}). The first deep   \suzaku\  (\citealt{Mitsuda07}) observation  performed in  2010 showed a spectrum of an obscured   AGN  ($\nhsym \sim 2 \times 10^{23}$\nh)  and,  surprisingly, two deep ($EW\sim 300$ eV)  blue-shifted absorption troughs at $E=7.4\pm 0.1$\,keV and $E=8.5\pm 0.2$\,keV  (\citealt{Braito2018}, hereafter B18).
  These are  most likely associated with  two zones of a highly ionized (log ($\xi/$\logxi)$\sim 5.5$) and high column density  ($\nhsym \sim 5-8 \times 10^{23} $\,\nh) outflowing wind with $v_{\rm out1}\sim -0.1\,c$ and $v_{\rm out2}\sim -0.2\,c$. 
The derived kinetic output ($\dot E_{\rm k}\sim 2.4\times 10^{44}$\,\lum)   is $\sim 8$\% of the bolometric luminosity  (or $\sim 2$\% of  $L_{\rm Edd}$), placing \sorg\ among the most powerful disk  winds.   A  deep follow-up  observation was  carried out in 2015 with   \xmm\ \& \nustar\    (of $\sim $ 130 ksec net exposure each).  This follow-up confirmed the presence of the  slow component of the wind, but did not confirm the $\sim 8.5$ keV feature previously observed in the \suzaku\ spectra. This observation unveiled also a possible faster ($v_{\rm {out}}\sim -0.35\,c$) component of the wind. Remarkably, during this   \xmm\ \& \nustar\ observation we    witnessed  an X-ray eclipse   caused by a denser ($ \Delta \nhsym \sim  10^{24}$\nh) streamline  of the wind moving across  our line of sight that   lasted $\Delta t \sim 120 $ ksec,  which is outflowing at $v_{\rm {out}}\sim -0.124\,c$.  A further investigation of the broad band X-ray emission, performed  adopting  a more self-consistent model for the pc-scale toroidal absorber (\textsc{mytorus}, \citealt{Murphy09})  confirmed that the  short-term X-ray spectral variability  cannot be accounted for by  $\nhsym$ variations of the neutral absorber, but certainly requires a variable highly ionized fast wind  (\citealt{Matzeu2019}, hereafter M19).

\sorg\ was also observed with ALMA, revealing a component that likely corresponds to a low velocity ($v_{\,\mathrm{CO}}\sim 170$ km s $^{-1}$) molecular outflow in the central  4\,kpc.  However,  as for PDS\,456 and  I\,Zw\,1 {\bf(\citealt{ReevesBraito2019})}, the kinetic power of the putative molecular outflow is at least two orders of magnitude below the expected value for an energy conserving wind ($\dot E_{\rm Mol}/\dot E_{\rm X}\sim 4\times 10^{-3}$; \citealt{Sirressi2019}). \\  Here we present the results of two  follow-up observational campaigns performed with \chandra\ in 2016 and with \swift\ in 2018, both of which caught \sorg\ in a  brighter state ($F_\mathrm{2-10\, keV}\sim 3.7 \times10^{-12}$\,erg\,cm$^{-2}$\,s$^{-1}$).
  We will show that the slowest component of the wind is persistent, as it is detected  in all the observations with a similar velocity of $v_ {\mathrm{out}}\sim - 0.07\, c$ and it is variable in column density, confirming its clumpy nature.  On the contrary   the second component of the wind  seen 
 during the \suzaku\ observation  appears more sporadic. 
 The paper is structured as follows: in \S 2 we briefly summarise the previous 
\suzaku, \xmm\ \& \nustar\ observations, which were already discussed in B18 and  in M19. The data reduction of  the \swift\ and \chandra\  observations is also presented in detail in  \S2.  In \S 3 we present  the spectral modelling focusing on the variability of the wind components. The discussion of the disk wind and  the implications of the long term variability are presented in \S 4.\ 
 Throughout the paper  we assume a concordance cosmology with  $H_0=70$  km s$^{-1}$ Mpc$^{-3}$, $\Omega_{\Lambda_{0}} =0.73$    and $\Omega_m$=0.27. For the abundances we used  those of \citet{Wilms2000}.\\

  \section{Observations and data reduction}
 \subsection{Earlier observations: \suzaku,   \XMM\ and \nustar}
  \sorg\ was observed in the X-ray band  in four epochs with five different satellites; in Table 1 we report the observation dates,  instruments and the net exposure times   for the all the  available observations.  
  
For the \suzaku, \xmm\ and \nustar\ observations  we adopt the same spectra as already described in detail in B18 and M19. As per the previous analysis,  the \xmm\ \& \nustar\ observations were split into two main intervals: slice\,A and slice\,B (see Fig.~5 of B18), where  slice A  is the relatively unobscured  half of the observation and slice\,B is where \sorg\ dropped in flux and appeared to go into a more obscured state (see B18 and M19 for details). 
For simplicity, here we  considered only the EPIC-pn spectrum as the agreement with the MOS spectrum is already discussed in M19. In this analysis  we combined the \nustar\ FPMA and FPMB spectra, because the individual detectors were consistent,  and   then grouped the resulting spectrum to reach at least 50 total counts per bin.  For these two epochs we will  limit our analysis to the best fits obtained in the previous works and simply test a new  grid of photoionised absorbers (see below).

   \subsection{\chandra}
\begin{figure} 
\begin{center}
\resizebox{0.49\textwidth}{!}{
\includegraphics[angle=0,width=0.62\textwidth] {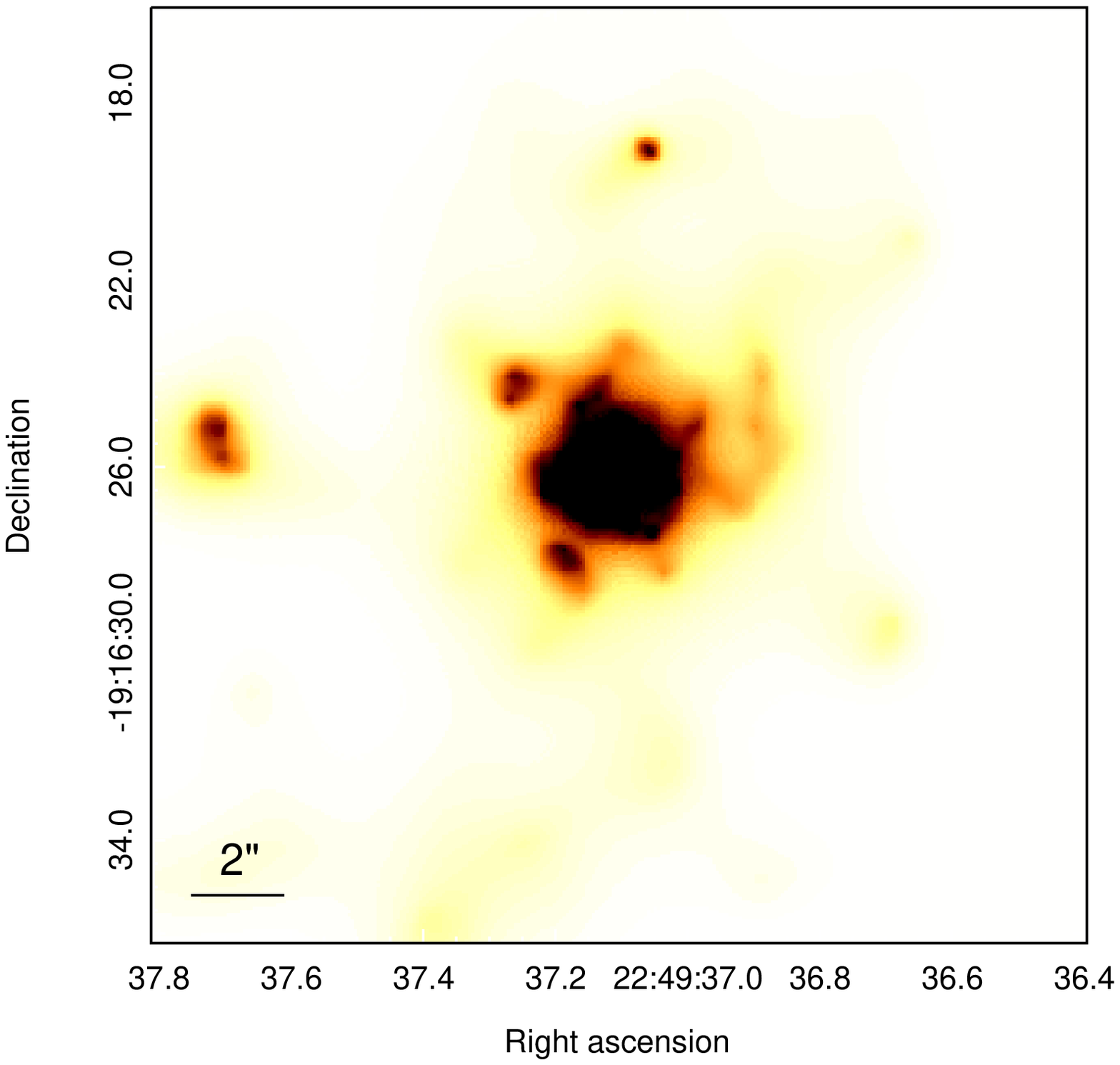}}
\resizebox{0.49\textwidth}{!}{
\includegraphics[angle=0,width=0.62\textwidth] {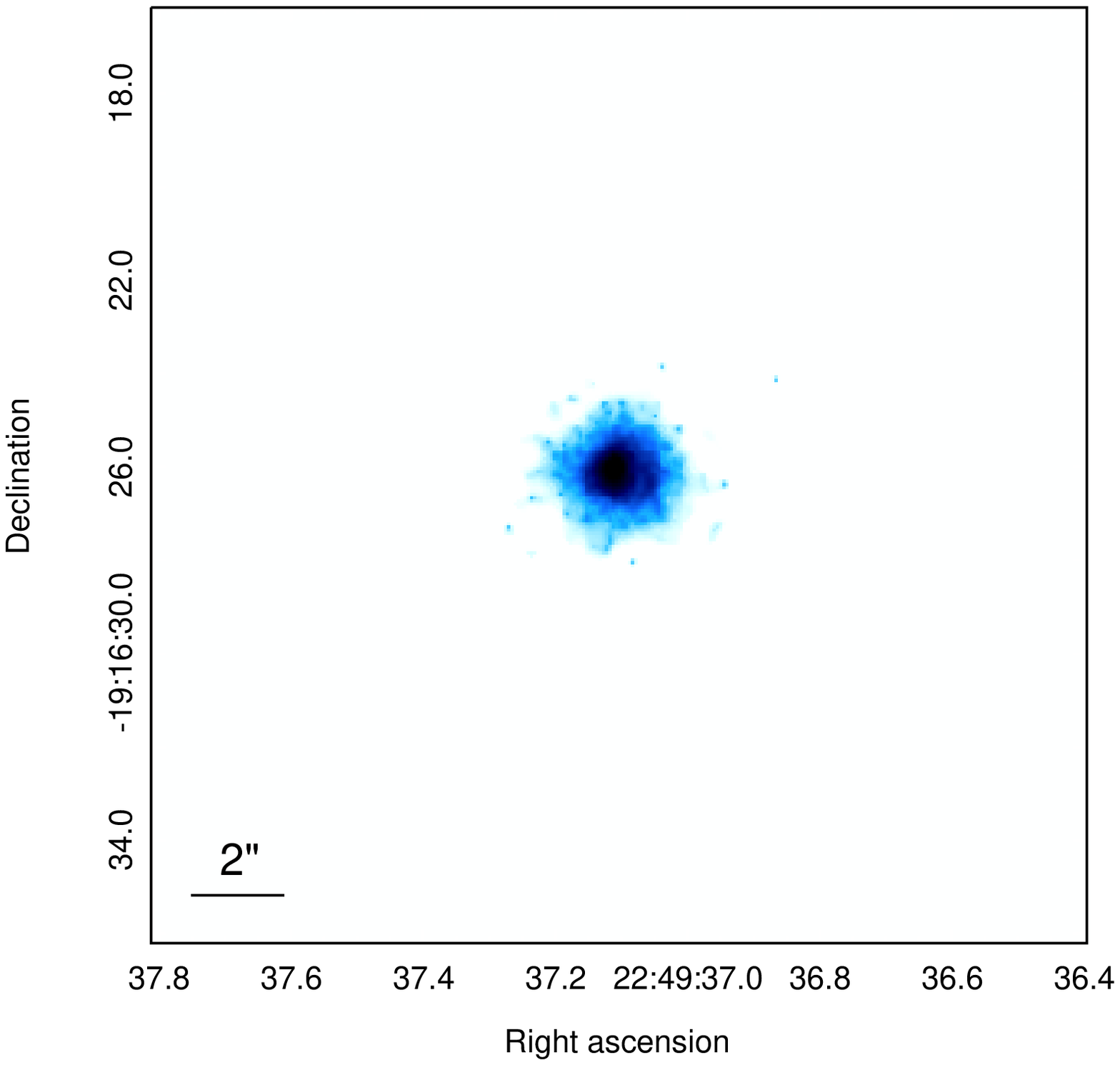}}
\caption{{ Upper Panel: \chandra\ soft (0.5-2 keV) image ($20''\times20''$) of \sorg.  The soft X-ray emission is mostly concentrated in the central  nuclear region (within 2$''$) with some weak  emission extending up to almost 4$''$.  The spatial scale is reported at the bottom right (note that for \sorg\ 1$''$ correspond to $\sim 650$ pc). Lower panel: hard  (3-8 keV) X-ray image  of \sorg. The hard X-ray emission is  close to a point like appearance  and originates from  the inner 1$''$. We used a   subpixeling factor of eight  to create the images, thus the pixel size is 0.062 arcsec. The images were  adaptively smoothed  with {\textit{dmimgadapt}}. }
 \label{chandra_images}
}
\end{center}
\end{figure}

 \sorg\ was observed     with \chandra\ in September 2016; the observation was split into two shorter exposures (see Table~\ref{table:log_observ}) that were carried out  five days apart.
The observations were performed with   the Advanced   CCD Imaging spectrometer (ACIS-S; \citealt{Garmire03})  in the 1/8 sub-array mode.  We reduced the data with the Chandra Interactive Analysis of Observation software (CIAO v. 4.10;  \citealt{Fruscione06})   and the latest Chandra Calibration Data Base  (CALDB version 4.8.1). We followed the standard reduction procedures and extracted source  and background spectra using a circular region of $2.5''$  and  $4.5''$ radius, respectively. We then  inspected the spectra extracted from the two exposures and verified that  there was no evidence for strong spectral  variability and only a moderate flux variation  between the two spectra ($\sim 15$\%).  We  thus added the spectra and combined the responses files with appropriate weighting.  The resulting spectrum was then binned  to  a minimum of 20 total counts per bin and modelled over  the $ 0.5 -9$\,keV  energy range.\\
We also extracted soft (0.5--2 keV)  and hard (2--8 keV) X-ray images. We first  corrected the absolute astrometry of the images adjusting the  aspect solution file and  the event file astrometry  with the CIAO tools \textit{wcs\_update} and \textit{reproject\_events}. We then merged the event files of the two observations using the CIAO tool \textit{dmerge}. Soft and hard X-ray images were  then created using a binning at $1/8$ of the native pixel size and then adaptively smoothed using  the tool {\textit{dmimgadapt}},  which assumes a gaussian function with sigma values ranging fro 1 to 30 image pixels   and a minimum convolution kernel of 2 counts\footnote{For details see: https://cxc.cfa.harvard.edu/ciao/ahelp/dmimgadapt.html}. The inspection of the images shows that  most of the X-ray emission originates in the innermost 1$''$ (or $\sim 650$\,pc), with only a weak residual soft X-ray emission extending up to $\sim 4''$ (see Fig.~\ref{chandra_images}). This is in agreement with the RGS spectral analysis presented in M19, where  the  soft X-ray emission was attributed to  the presence of gas that is mainly photoionised by the central AGN and possibly associated with the emission of the circumnuclear Narrow Line Region gas.

\subsection{\swift}
 \swift\ observed \sorg\  in 2018 for a total exposure time of $\sim$ 70 ksec. The observational campaign consisted of six XRT exposures in the standard PC-mode, with three long ($\sim 19$\,  ksec each) exposures and three shorter  observations (see Table~1).  The observations were performed as  a monitoring program aimed  at investigating the possible variability of the neutral absorber on time scales as short as a day. The observations were  performed in April 2018  covering about 10 days.   \\
 We extracted the  source and background spectra from each of the observations adopting a circular region with a radius of 42$''$ and   78$''$, respectively.  We also extracted broadband as well as soft (0.3--2 keV) and hard  (2--10 keV) X-ray light curves  and found no evidence for spectral variability and only minor flux variations within  $\sim 20$\%.  The inspection of the spectra extracted from all the observations suggests that \sorg\ was caught  in a similar state in each of them. Before adding all the six  spectra, we fitted all of them  with a baseline continuum model composed of: an absorbed power component,  a scattered power-law emission and a thermal emission component, finding consistent spectral parameters and only moderate flux variations.
  We also note that the residuals of all  three longest observations  show a drop above  7 keV.  However,  the  low count statistics  in the individual spectra prevent us from investigating the nature of the drop  in the  single observations,  therefore we  combined all of them into a single spectrum.  We then binned the combined spectrum to a minimum of 20 total counts per bin and  considered the $0.3-9$\,keV  energy range  for the following spectral analysis.

   \begin{figure} 
\begin{center}
\resizebox{0.49\textwidth}{!}{
\rotatebox{-90}{
\includegraphics{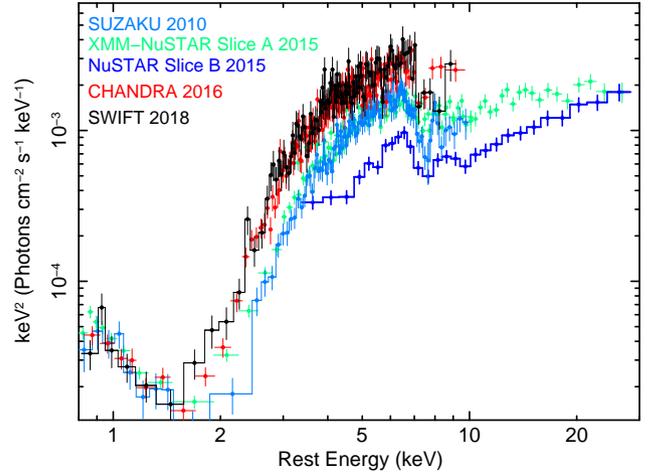}
}}
\caption{{Fluxed spectra of all   the recent X-ray observations of \sorg\ showing the  short (on timescale of a day) and long term variability of the X-ray emission. The \suzaku\ spectrum (collected in 2010) is shown in light blue. The \xmm\ \& \nustar\ simultaneous observation   performed in 2015 (slice\,A) are shown in  green. The  second half of the \nustar\ observation (slice\,B), where the eclipsing event occurred, is shown in blue.   The  spectra collected during the most recent observations performed with \chandra\ in 2016 and \swift\ in 2018 (average) are shown in red and black, respectively.  To generate the plot, we created fluxed spectra by unfolding the data against a $\Gamma=2$ power law model. The data have been rebinned for plotting purposes.}
 \label{fig:all_gamma2}
}
\end{center}
\end{figure}

   \section{Spectral analysis}
   \subsection{Baseline model}
   
 In Fig.~\ref{fig:all_gamma2} we show  the fluxed spectra of all the X-ray observations, obtained by unfolding the data against a power law model with $\Gamma=2$.  Two facts emerge by comparing all the observations.  First of all,  the spectral curvature, that we can ascribe to the presence of the neutral absorber,  appears to be similar in all the spectra,   with the exception of the second part of the \nustar\ observation (slice\,B, see Fig.~\ref{fig:all_gamma2}, dark blue spectrum). Second and more importantly   all the spectra show  a drop around $7.2-7.4$ keV. 
We also note that in the two  new observational campaigns (Fig.~\ref{fig:all_gamma2}, red and black data points for the \chandra\ and \swift\ spectra, respectively) we caught \sorg\ in a relatively brighter state ($F_{\rm{(2-10)\,keV}}\sim 3.7\times 10^{-12}$\,\flux; see Table 2). \\

 The best fit model found for the first two epochs  (e. g. B18) was  defined as: 
   
   \begin{equation*}
   \begin{split}
 F(E)=\textsc{tbabs}\times[\textsc{zpow}_{\rm scatt}+\,\textsc{xstar}_{\rm em}+\,\textsc{mekal}+  \,\textsc{pexmon} \\
 +\,\textsc{zphabs}\times \textsc{xstar}_{\rm FeK,1}\times\textsc{xstar}_{\rm FeK,2}\times\textsc{zpow}]
\end{split}
\end{equation*}
Here  \textsc{tbabs} represents the neutral Galactic absorption ($\nhsym=2.5\times 10^{20}$\,\nh; \citealt{Dickey}),  \textsc{zpow}$_{\rm scatt}$ is the scattered power-law component; the cold reflection component is modelled with \textsc{pexmon}  (\citealt{pexmon}), which includes also the   \feka,  \fekb\ and the Ni $K\alpha$ emission lines as well as the \feka\ Compton shoulder.   We fixed the inclination angle to 60 degrees and the high energy cutoff to 100 keV. We fixed also the amount of reflection $R=\Omega/2\pi=1$, while the normalisation is  allowed to vary.  For   the   reflection component we tied its photon index  $\Gamma$  to the primary power-law component, while we allow the scattered power-law component to have a different slope to account for any soft excess. For the soft X-ray emission  the best fit models include both the  emission from  a  collisionally ionized plasma (\textsc{mekal} component, with $kT=0.84\pm 0.08$\,keV) and the optically thin gas photoionized by the AGN (\textsc{xstar}$_{\rm em}$;  see M19). Finally,  \textsc{xstar}$_{\rm FeK,1}$ and  \textsc{xstar}$_{\rm FeK,2}$ are the two photoionised absorbers. \\ 

\begin{figure} 
\begin{center}
\resizebox{0.48\textwidth}{!}{
\rotatebox{-90}{
\includegraphics{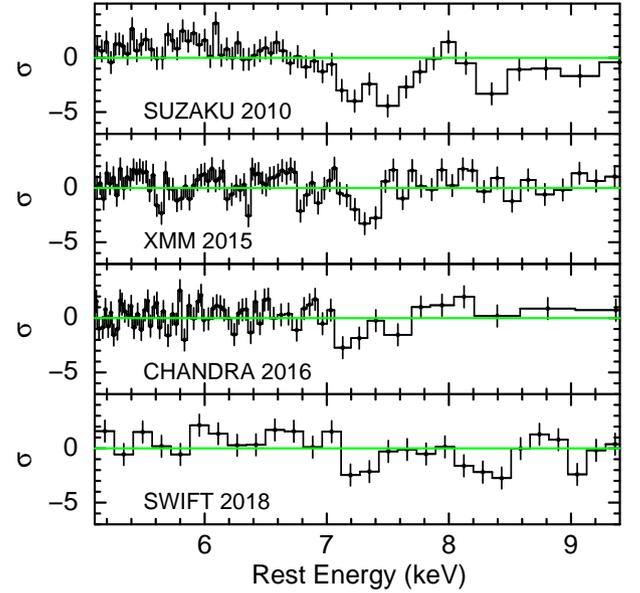}
}}
\caption{{Residuals for all the observations  to the continuum model  where we  assumed the same photon index for the primary power law component. The continuum model   is composed  by a soft power-law component, a collisionally and a photoionized  plasma emission components and a primary  power-law component transmitted through a  neutral  absorber.   For the 2015  \xmm\ \& \nustar\ observation we include also a reflection component modelled with {\textsc pexmon} (\citealt{pexmon}).  The $\sim 7.4 $\,keV absorption trough is at its weakest during the \XMM-slice\,A. It is also apparent that, while a higher energy absorption trough is present at $\sim 8.5$ \ keV  in the 2010  and 2018  spectra  it is undetected in the 2015 \XMM\  and  2016 \chandra\ observations. The \swift\ data are binned adopting a constant energy bin of 150 eV for plotting purposes.  }  \label{fig:residuals}
}
\end{center}
\end{figure}

  We first proceeded to fit  the  new \chandra\ and \swift\ spectra with  the best fit continuum model found for the previous observations, minus the  two photoionised  absorbers.   In Fig.~\ref{fig:residuals} we show the residuals  in the iron absorption region of all the spectra to the baseline continuum model,  where for completeness  we also report  the residuals of the  previous \suzaku\ and \xmm\ \& \nustar\ spectra. For simplicity for the 2015 observation,  here  we only show the \XMM-pn data of slice\,A.  A  similar comparison between slice\,A and the slice\,B residuals is shown in Fig.~ 8 of B18.  We note that for the \xmm\ \& \nustar\ spectra (both slice\,A and slice\,B) we  included a neutral reflection component to the baseline continuum model (as above), which we modelled  with the  \textsc{pexmon} component allowing only   its normalisation to vary. For the \chandra\ \& \swift\  datasets, which lack the bandpass above 10 keV, this is not required.   Two main facts emerge when inspecting the residuals.  First of all   a trough at $\sim 7.2-7.6 $ keV  appears to be  present in all the  observations, albeit it may vary in depth; as we investigate below this may be ascribed to  the  variable lower velocity zone of the disk wind. The second fact that emerges is that  a higher energy absorption structure is seen at $\sim 8.5 $ keV  in only two of the four epochs,  the \suzaku\ and  \swift\ spectra.\\

 We then  reapplied  the best fit  model, which includes two  multiplicative grids of  ionised absorption models,  to the previous \suzaku, \xmm\ \& \nustar\ spectra   (both slice\,A and slice\,B) and also to the new \chandra\ and \swift\ observations. 
We considered each observation separately, with the exception of the \chandra\ \& \swift\ observations, which caught \sorg\  with a similar continuum and  flux level, where we tied only the main parameters of the continuum, but we allowed the $\nhsym$ of two ionised absorbers to be independent from each other. We assumed for all the observations the  same photon index as currently  measured  with \nustar\ ($\Gamma=2.17\pm 0.06$, see Table~2), because   we lack  any coverage above 10 keV, where the primary continuum emission emerges.  We note that the $\Gamma$ derived here is marginally different from the one reported by M19 for a similar configuration (Model A in M19: $\Gamma=2.25 \pm 0.07$); however it is well within the errors. Similarly to what was found by M19, when the photon index  of the soft scattered component is allowed to vary it  tends to a high value ($\Gamma_{\rm{ soft}}=3.7\pm 0.3$). However,  the precise modelling of the soft X-ray  emission does not affect the parameters of the neutral absorber or  of the disk wind (see also M19). \\

  A difference of the fits presented here is that to model the two zones disk wind  we  generated a new  grid of photoionised absorbers with  the \textsc{xstar} photoionisation code (\citealt{xstar}). The  grid was generated assuming a high turbulence velocity  ($v_{\rm turb}=10^{4}$\,km s$^{-1}$). We assumed  a SED similar to IZW1, where the soft (between 0.3 and 1.2 keV) photon index ($\Gamma$) is $\sim 3$   and the hard X-ray $\Gamma$  is 2.2 (see Reeves \& Braito 2019). This is justified by the fit reported by M19, where the soft X-ray slope was indeed of the order of 3,  in agreement with the slope measured here (see above).    The grid was optimised for high column density  ($\nhsym=10^{22}$  - $2\times10^{24}$\nh) and high ionisation (log($\,\xi /{\rm erg\,cm \,s^{-1})}= 3 - 7$) absorbers.  The main difference to the grid  used in the previous works is that this new grid has  finer steps\footnote{The  xstar grid was generated  with 20 linear steps for  $\xi$  and 10  logaritmic steps for $\nhsym$.}  (i. e.  it interpolates between   a larger number of ionized absorbers). The choice of the high turbulence velocity is  justified by the broadening measured for the absorption features. For instance, if we model the two absorption features seen in the \suzaku\ spectra with two Gaussian absorption lines, we measure $\sigma \sim 0.35\pm 0.08$\,keV (or $\sigma\sim 1.4\times 10^4$ km s$^{-1}$, see also B18). 
Nevertheless, we also tested   grids  with a lower turbulence velocity  (i. e. $v_{\rm turb}=1-5 \times 10^{3}$\,km s$^{-1}$), which resulted in worse fits. In particular, for $v_{\rm turb}=1000$\,km s$^{-1}$ the fit of the \suzaku\ spectra  is worse by  $\Delta \chi^2=31.4$;  the predicted absorption features are too narrow and  cannot reproduce the breadth   and depth  of the observed profiles.   Note that,  as discussed in B18, the equivalent widths of  the absorption lines seen in the \suzaku\ spectra is of the order of $\sim 300$\,eV.
 Another difference with the previous fits is that here we assumed  the ionisation of the two zones to be the same.
 
   In Table~\ref{tab:bestfits},  we report the new parameters obtained for these observations, using the new grid of models.

\subsection{Slower outflow component}
 Given the known degeneracy between   $ \nhsym$ and log\,$\xi$, we proceeded to investigate the variability of the Fe K absorption troughs by allowing only the  column density to vary and assumed a constant ionisation (see below). We also allowed the outflow velocity  to vary   between the observations. \\
 In Table~\ref{tab:bestfits} we summarise the results of the best fits for each of the datasets. The inclusion of the first zone of the wind improves all the fits for a $\Delta \chi^2/\Delta \nu=107.5/3$, $\Delta \chi^2/ \Delta \nu=27.2/3$ and $\Delta \chi^2/\Delta \nu=17.3/3$  for the \suzaku, slice\,A \& slice\,B  and  the \chandra\ \& \swift\ data, respectively. 
 
\begin{figure} 
\rotatebox{-90}{
\includegraphics[angle=0,width=0.3\textwidth]{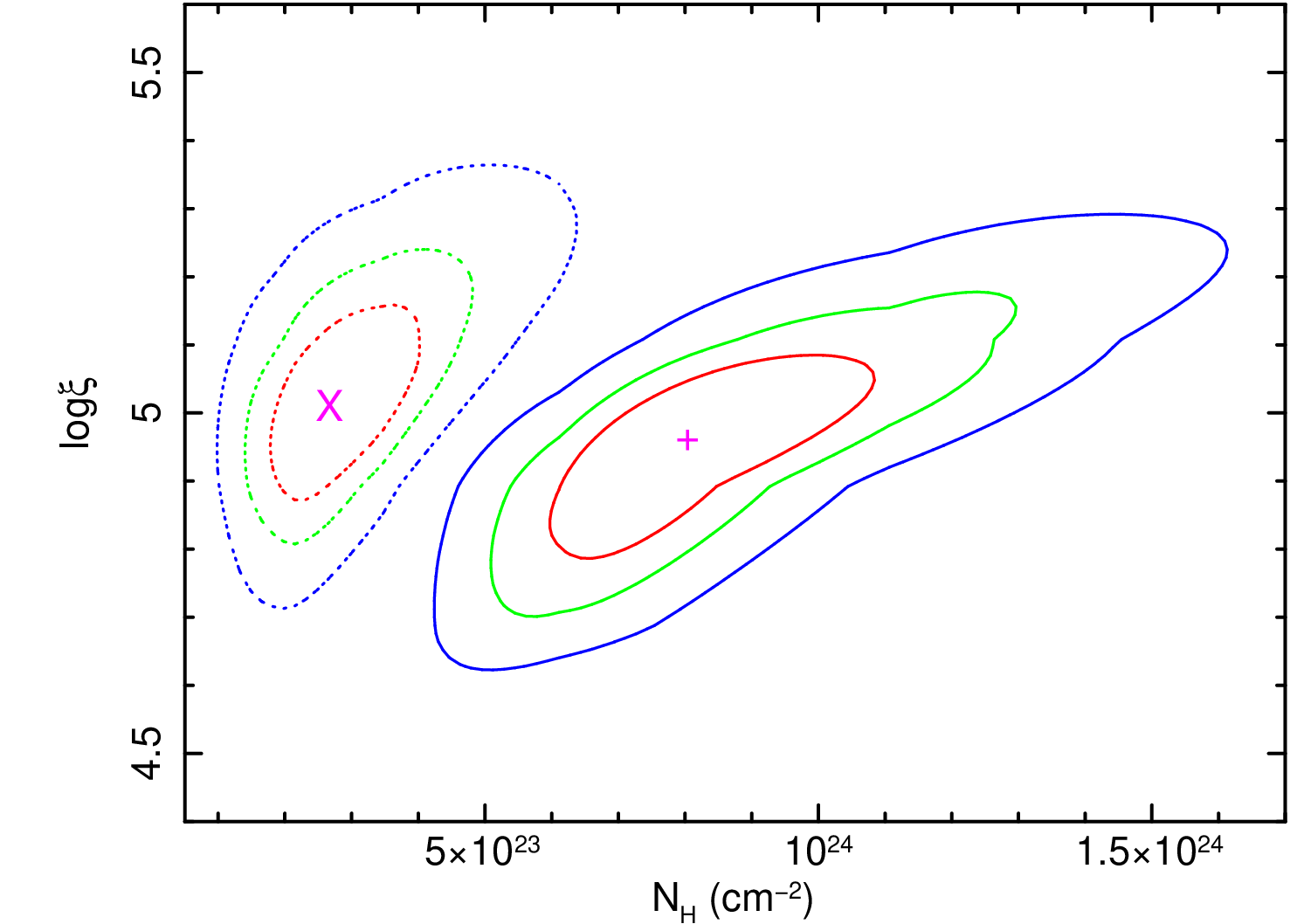}
\includegraphics[angle=0,width=0.3\textwidth]{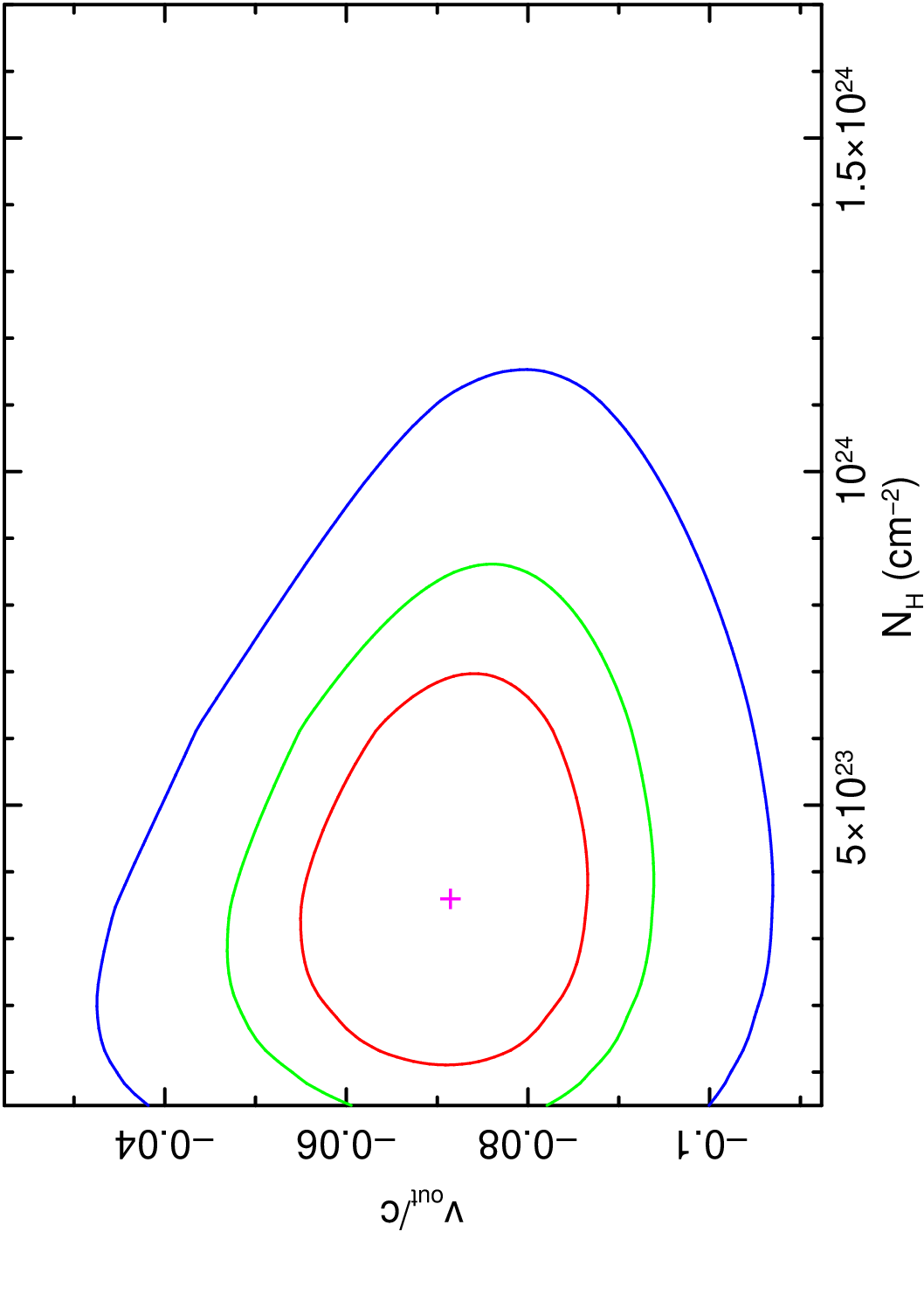}
}
\caption{{Upper Panel: contour plots  for the ionisation against the $\nhsym$   for  the slow component of the disk wind in the \suzaku\ (solid lines) and the \xmm\ \& \nustar\ slice A spectra (dotted lines).   This clearly demonstrates the variability of this zone in the log$\xi$-$\nhsym$ plane.  Lower Panel:  contour plots  for the disk wind velocity against the $\nhsym$   for the \swift\ observation.  The    red, green and blue  lines correspond to  the 68\%, 90\% and 99\%  confidence levels for two interesting parameters. }\label{fig:cont1}
}
 \label{fig:cont}
\end{figure}

 We found that, with the exception of  the obscured slice\,B,   the outflow velocity is consistent within errors over  all the epochs,  ranging from $v_{\mathrm {out1}}/c\sim-0.072$ to $v_{\mathrm {out1}}/c\sim-0.077$, while  the column density of this lower velocity zone varies among all of the observations.  The wind opacity is at its maximum level during the \suzaku\ observation in 2010 ($\nhsym= 8.1^{+3.0}_{-2.4}\times 10^{23}$\,\nh; see Table~2) and it is at its lowest level  in slice\,A  with a measured $\nhsym=2.6^{+1.1}_{-0.9} \times 10^{23}$\,\nh. Despite this large variation in column density, the fluxes of these two epochs are identical. This likely rules out changes where the absorber is only reacting to the continuum, via changes in ionisation and implies that the absorber variations are intrinsic  and due to the column density.
Finally we investigated  if, in our  best-fit models,  there is any  physical degeneracy  between  the column density, the ionisation or the outflow velocity.  We therefore consider the \suzaku\ and  the non-occulted part of the \xmm\ \& \nustar\ observations and allowed also the ionisation to be independent. In Figure~\ref{fig:cont} (upper panel) we show the   confidence contours obtained for  the ionisation versus the $\nhsym$  for both  these observations. Although the  elongated shape of the contours suggests  some degeneracy between    log\,$\xi$ and $\nhsym$, it is clear that the inferred variability of the opacity of this zone of the wind does not depend on the  assumption of a constant ionisation. For the \chandra\ and \swift\ observations, since the ionisation is poorly constrained, we investigated the dependence of the $\nhsym$ and the wind velocity. The relative contours  for the \swift\   spectrum are shown in Figure~\ref{fig:cont} (lower panel). 
 
\subsection{Faster outflow components} 
\begin{figure} 
\begin{center}
\resizebox{0.49\textwidth}{!}{
\rotatebox{-90}{
\includegraphics{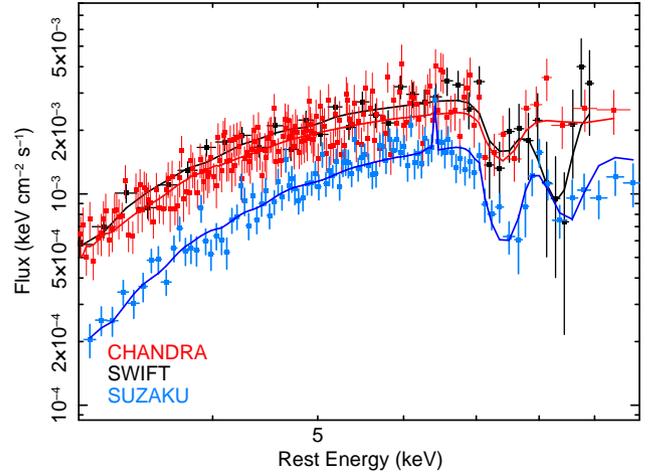}
}}
\caption{{ Best fit model and spectra  for the   \swift\ (black data points), \chandra\ (red data points) and \suzaku\  (light blue) observations. The  model includes two ionized and outflowing absorbers, for which   we assumed  the same ionisation, but we allow the column densities to vary.  Note that  a deep   absorption trough    at $\sim 7.3  $ keV is seen in  all the observations, while the second and deeper ($EW\sim 300$ eV) structure at $\sim 8.5$ keV is detected only in the \swift\  and \suzaku\ observation. The \swift\  data were rebinned to adopting a constant energy binning of $150 $ eV for plotting purposes. } \label{fig:bestfitCH_SW}
}
\end{center}
\end{figure}

Regarding the second  fast zone, responsible for the  8.5 keV  absorption feature, the  fit  substantially improves only for the \suzaku\  ($\Delta \chi^2/\Delta \nu=35.0/2$) and \swift\ ($\Delta \chi^2/\Delta \nu=10.0/2$) observations.  Note that in both cases the fit improvement indicates that this zone is required at a  $> 99.9$ \% confidence level. There is no evidence for this zone in the \chandra\  spectrum, where we are able to derive a rather stringent  upper limit   for the $\nhsym$ of this faster zone of $\nhsym <1.5\times 10^{23}$\,\nh. 
 We found  that the  column density  and outflow velocity ($v_{\mathrm {out2}}/c\sim-0.2$; see Table~2) of this zone are rather similar between the 2010 and 2018 observations, suggesting that our line of sight intercepts a similar inner zone in both observing periods.
In Fig.~\ref{fig:bestfitCH_SW} we compare the \suzaku, \chandra\ and \swift\ spectra and the best fit models.  It is noticeable that  during the \swift\ observation  the faster zone is similar to the fast zone detected in 2010.  It is also clear that such a zone is not present during the \chandra\ observation, despite that it caught \sorg\ at a similar flux level as seen in the 2018 \swift\ campaign. 

As already discussed in B18 and M19,  an even faster zone could be present during the 2015 \nustar\ observation  ($\Delta \chi^2/\Delta \nu=36.2/2$) with an outflowing velocity of $v_{\mathrm {out3}}/c=-0.35\pm0.01$ and an $\nhsym =(3.8\pm 1.2)\times 10^{23}$\,\nh.  In contrast to  what was assumed in the previous works,   here the ionisation of this zone was tied to the ionisation of the slow component.  We note that  the assumed value is within the errors  of  the log$\xi$ previously reported. The imprint of this zone is the absorption structure seen in the \nustar\  spectra at around 10 keV  (see Fig.~8 of B18); thereby  we cannot assess if it is a persistent or sporadic streamline as  it cannot be detected in the other observations, which lack the higher energy bandpass required to detect it.

\begin{table*} 
  \caption{Summary of the   two phase disk wind model applied to all the observations.  
$^a$: the normalisation units are   $10^{-3}$ ph keV$^{-1}$\,cm$^{-2}$.
$^b$: The \xmm\ \& \nustar\ 2015 slice\,A and slice\,B spectra were fitted simultaneously.
$^c$ : We perfomed a joint fit of the \chandra\ and \swift\  spectra,   because \sorg\ was in a similar flux state.
$^t$: denotes parameter was tied.
$^f$: denotes that the parameter was fixed. The fluxes are corrected  only for the Galactic absorption, while the luminosities are intrinsic. \label{tab:2zones_nustar}
 }
   \begin{tabular}{llccccc}
\hline
 Model Component  &  Parameter  &  \suzaku & Slice\,A & Slice\,B  &\chandra  & \swift \\ 
 \hline
&&   \\
Primary Power-law &$\Gamma$ & 2.17$^f$ & $2.17_{-0.06}^{+0.06}$  & $2.17{^t}$ & 2.17$^f$ & 2.17$^f$\\
& Norm.$^a$ & $3.1_{-0.2}^{+0.2}$&  $2.6_{-0.3}^{+0.3}$ & $1.5_{-0.2}^{+0.2}$ & $4.4\errUD{0.2}{0.2}$ & $4.9\errUD{0.3}{0.3} $\\

&&&&&\\
Neutral absorber &$N_\mathrm{H}(\times 10^{23}$ \nh)& $2.7\errUD{0.1}{0.1}$ & $2.2\errUD{0.1}{0.1}$  &$2.2^t$ & $2.0\errUD{0.1}{0.1}$ & $1.9\errUD{0.1}{0.1}$ \\
 &&\\
Zone 1     & $N_\mathrm{H1 } $($\times 10^{23}$ \nh)&$8.1\errUD{3.0}{2.4}$ &  $2.6\errUD{1.1}{0.9}$& $5.0\errUD{2.2}{1.3}$  & $3.6\errUD{2.4}{2.1}$ &$3.7\errUD{3.6}{2.8}$ \\
              &log$ \xi_1$&  $5.0\errUD{0.1}{0.2}$ &  $5.0^{f}$ & $3.70\errUD{0.11}{0.16}$  &$5.0^{f}$&  $5.0^{f}$\\
               &$v_\mathrm{out1}/c$&$-0.077\errUD{0.009}{0.009}$ & $-0.073\errUD{0.022}{0.022}$  &$-0.13\errUD{0.02}{0.02}$  &$-0.072\errUD{0.017}{0.017}$&$-0.072^t$ \\
&  $\Delta \chi^2/ \Delta \nu$    & 107.5/3&\multicolumn{2}{c}{$50.6/5^b$ }&  \multicolumn{2}{c} {$17.3/3^c$}\\

&&&&&\\
Zone 2    & $N_\mathrm{H2 }$ ($\times 10^{23}$ \nh)&$6.0 \errUD{3.1}{2.2}$ &  $3.8\errUD{1.2}{1.2}$ & $3.8^t$ &$<1.5$&$7.0 \errUD{5.6}{4.1}$ \\
              &log$ \xi_2$&$5.0^t$ & $5.0^t$ & $5.0^t$&$5.0^t$ & $5.0^t$  \\
&$v_\mathrm{out2}/c$ & $-0.20\errUD{0.02}{0.02}$ & $-0.35\errUD{0.01}{0.01}$ &$-0.35^t$  &$-0.17^{f}$&$-0.17\errUD{0.03}{0.03}$\\
&  $\Delta \chi^2/\Delta \nu$    & 35.0/2&\multicolumn{2}{c}{$ 36.2/2^b$ }&  \multicolumn{2}{c} {$10.0/2^c$}\\

&&&&&\\
 &$F_{(2-10)\,\mathrm {keV}}\times 10^{12}$ (\flux) &   2.0 &  $2.1$ & 1.0&  $3.8$&3.6\\

 &$L_{(2-10)\,\mathrm {keV}}\times 10^{43}$ (\lum) &   1.4 &  $1.2$ & 0.8&  $2.0$&2.2\\
&&&&&\\
  $\chi^2/\nu$ &  & 230.1/202 &\multicolumn{2}{c}{$468.2/395^b$ }&  \multicolumn{2}{c} {$339.7/340^c$}\\
               
     \\
   \hline
    \label{tab:bestfits}
\end{tabular}
\end{table*}

   \section{Discussion}
  We have presented the analysis of the current X-ray observations of the disk wind in \sorg.  Here, multiple  and variable wind components with velocities ranging from $\sim -0.08\,c$ to $\sim- 0.2\,c$   (and potentially up to  $0.35\,c$) are seen at different times. Multi epoch  observations  of disk winds, like the one presented here, are crucial for revealing all the possible phases of the disk wind. For example, over a decade worth of observations of  PDS\,456  revealed that the wind is most likely clumpy and/or stratified  with the ionisation ranging from log\,($\xi /\rm{erg\, cm \, s^{-1})}\sim 2$\, \logxi\   up to log\,($\xi /\rm{erg\, cm \, s^{-1})}=6$\,\logxi\     and velocities ranging from $\sim -0.2\,c$ up to $\sim -0.46\,c$ ({\citealt{Reeves2016,Reeves2018,Reeves2020}).  It is possible, as suggested in other examples of ultra fast disk winds,  that we are looking at a stratified wind, where multiple components are launched at different disk radii, but not all of them are always detected.   This adds \sorg\ to the  small but growing list of multiphase fast X-ray winds.  Other examples of AGN with at least two variable phases of the X-ray winds are PG\,1211+143 (\citealt{Pounds2016,pg1211}),  IRAS  13224-3809 (\citealt{Chartas2018,Parker2017,Pinto2018}),  1H\,0707-495 (\citealt{Kosec2018}), IRAS 17020+4544 (\citealt{Longinotti2015}5) and PG 1114+445 (\citealt{Serafinelli2019}). In those cases, multiple phases with a common or different outflowing velocities are detected in the X-ray band.  In contrast to most of the cases reported so far,  neither of  the two phases seen in \sorg\   requires a different ionisation (aside from slice\,B) suggesting that we are seeing different streamlines of the same highly ionized flow. The only exception could be the eclipsing event seen in 2015, where a solution is found with  a  lower  ionisation  for the Fe K intervening absorber.   However, what we most likely see during this occultation event is a higher density and lower ionisation  clump of the wind, which  could be faster because   its higher opacity makes it  easier to accelerate  (\citealt{Waters2017}). Note that this does not imply that the soft X-ray wind components, like the ones seen for example in PDS\,456 or   PG\,1211+143,  are not present;  in contrast to the other examples, \sorg\  is seen through  a relatively high column density  ($\nhsym\sim 2\times 10^{23}$\, \nh, see Table 2) neutral  absorber, therefore these phases may  be  hidden behind it.    \sorg\ is not the only example where multiple Fe-K zones  with the same ionisation and outflowing with different velocities had been detected in a single observation. For instance, two simultaneous Fe-K phases were detected at least twice in PDS456 (\citealt{Reeves2018,Reeves2020}) and possibly in PG\,1211+143 (\citealt{Pounds2016})  and  IRAS\,13349+2438 (\citealt{Parker2020}).    
  
  We now discuss the  properties of the  various phases of the disk wind and the implications for the overall energetics. Evidently, if we want to understand the possible impact of the disk wind on the host galaxy, it is important not only to estimate  the mass outflow rate and kinetic energy that  the outflow can transport at  a specific epoch, but also to account for all the phases of the wind and their variability. Crucially, what we need to establish is whether or not the wind is persistent  and if variable, its average   kinetic energy.  Overall, the wind detected in \sorg\ is at least composed by two variable zones: a persistent slow component with an outflow velocity of $v_{\mathrm{out1}}\sim -0.08\,c$ and a sporadic second faster zone with a velocity of $v_{\mathrm{out2}}\sim -0.2\,c$.
  
  \subsection{The slower outflow component}
  For the estimate of the outflow rates,  we first  concentrate on the lowest velocity zone, because it is clear that this  phase  (zone 1) is persistent as it is  detected in all the observations analysed so far.  We note that this zone  shows an extreme variability with both long  and short term variations of the  $\nhsym$,  suggesting that the wind is  inhomogeneous.  While the $\nhsym$ of this zone varies among the  different epochs, the velocity appears to be constant within the errors (see Table~2, slice\,B excepted). In particular,  while the hard X-ray luminosity varies by a factor of $\sim 2.8$ (see Table~2), the wind does not appear to directly respond to it   in either velocity or ionization. 
 Our interpretation is that we are most likely seeing  the same component, which is always present   but clumpy as indeed we have already witnessed an occultation event,  when during slice\,B our line of sight intercepted a higher opacity clump of the wind (see Table~\ref{tab:bestfits}).\\
   
The crucial parameter needed  to infer the energetics of the disk wind is its mass outflow rate $dM/dt$ ($\dot M_{\rm out}$). This can be derived using the equation (see \citealt{NardiniScience}): 
 \begin{equation}
 \dot M_{\rm {out}}=\Omega \,  \mu \, m_{\mathrm p}\,\, v_{\rm {out}}\, R_{\rm w}\, N_{\rm {H}}
\end{equation}

   where $\mu$ is  a constant factor  set to $\mu=n_{\rm H}/n_{\rm e}=1.2$ for solar abundances, $\Omega$ is the wind solid angle, $R_{\rm{ w}}$ is the disk wind radius,  $\nhsym$  and  $v_{\rm {out}}$ are the column density and the velocity of the disk wind. \\
Since  we do not know the exact geometry  for \sorg, following the same argument  presented in \citet{Gofford2015}   and \citet{Tombesi2013}, we assumed that  the wind subtends  $\Omega/4\pi = 0.5$.} Indeed,  the systematic search of  fast disk winds   in  bright nearby local AGN resulted in a detection rate of   about 40\%, suggesting these  winds have a wide opening angle. Note that   a large  solid  angle of about $2 \pi$ was derived for the wind in PDS~456 from  its  P-Cygni   Fe K profile     (\citealt{NardiniScience}).  \\
The  main parameter  that we now need to quantify  is the launching  radius of the wind; a lower limit of the radial distance can be derived from the outflow velocity   assuming that  the wind is launched at  its escape radius $R_{\mathrm{min}} = 2\,G\,M_{\rm {BH}}/v^2$.  We note that by adopting this radius, we derive the most conservative estimate of the mass outflow rate and energetics (see \citealt{Gofford2013,Tombesi2012}). 
 As the main  uncertainty  in this latter equation is the black hole mass of \sorg, we decided to normalise the mass outflow rate  to the Eddington rate:

\begin{equation}
 \dot M_{\rm {Edd}}= 4\, \pi\, G\, m_{\mathrm p}\, M_{\rm {BH}}\, /\,\sigma_{\rm {T}}\,\eta\, c
  \end{equation} 
  
\noindent where $\sigma_{\rm {T}}$ is the Thomson cross section and $\eta=0.1$ is the   accretion efficiency. 
Thus combining equation (1) and (2)  and substituting for $R_{\rm{min}}$ we obtain:
\begin{equation}
\dot M=\frac{\dot M_{\rm {out}}}{ \dot M_{\rm {Edd}}}=2 \frac{\Omega}{4\pi} \mu \,N_{\rm {H}}\,\sigma_{\rm {T}}\,\eta\,\left(\frac{v_{\rm{out}}}{c}\right)^{-1}
    \end{equation}

The kinetic power ($ L_{\rm {KIN}}=1/2 \dot  M_{\rm {out}}\, v_{\rm{out}}^2$) of the wind can also be  normalised to the Eddington luminosity  ($L_{\rm{Edd}}
=\eta\, \dot M_{\rm {Edd}} \,c^2$). Thus  substituting for $\dot  M_{\rm {out}}$ and  $\dot M_{\rm {Edd}}  $ the wind kinetic power is:
\begin{equation}
\dot E= \frac{L_{\rm {KIN}}}{L_{\rm {Edd}}}=\frac{\Omega}{4\pi}\, \mu\, N_{\rm {H}}\,\sigma_{\rm {T}}\frac{v_{\rm{out}}}{c}
    \end{equation} \\

\begin{table*} 
  \caption{Summary of the energetics of the different phases of the disk wind. 
$^a$:  Mass outflow rate ($\dot M$)  in Eddington units.
$^b$:  Here zone 2 refers to the faster zones seen outflowing at $v\sim -0.2 c$ in the \suzaku\ and \swift\ observations and at $v\sim -0.35\,c$ in the \XMM-\nustar\ observation, respectively.
$^c$: Outflow kinetic power ($\dot E$) in Eddington units
$^d$:  The kinetic power unit  is  \lum and is derived for $M_{\rm{BH}}=10^8\, M\odot$.
$^e$:  The outflow   energetics   for the molecular gas depends   on  the CO-to-H$_2$ conversion factor, here a factor of $2.1 M_\odot$ (K km s${-1}$ pc$2$)$^{-1}$ was adopted (Sirressi et al. 2019).
  \label{tab:energtics}
 }
   \begin{tabular}{llll}
\hline
 Observation & parameter & zone 1$^a$ &zone $2^b$ \\
 \hline
&& &  \\
\suzaku\ 2010 & $\dot M$ &$0.84\errUD{0.33}{0.27}$&$0.24\errUD{0.13}{0.09}$\\
&$\dot E^c$ &$0.025\errUD{0.01}{0.008}$&$0.048\errUD{0.025}{0.018}$\\
 &log\,$L^d_{\mathrm{KIN}}$&44.5&$44.8$\\
& $\dot p_{\mathrm {w}}$ ($L_{\mathrm{bol}}/c$) & 2.7&2.0  \\
&& &  \\
\XMM-\nustar 2015 slice\,A  & $\dot M$ &$0.28\errUD{0.15}{0.12}$&$0.09\errUD{0.03}{0.03}$\\
&$\dot E$&$0.008\errUD{0.004}{0.004}$&$0.053\errUD{0.017}{0.017}$\\
 &log\,$L_{\mathrm{KIN}}$&44.0&$44.8$\\

&$\dot p_{\mathrm {w}}$ ($L_{\mathrm{bol}}/c$) &  0.9&1.0\\
&& &  \\
 && &  \\
\chandra\ 2016 & $\dot M$ &$0.40\errUD{0.28}{0.25}$&$<0.07$\\
 & $\dot E$&$0.010\errUD{0.007}{0.006}$ &$<0.01$\\
 &log\,$L_{\mathrm{KIN}}$&44.1&$<44$\\
&$\dot p_{\mathrm {w}}$ ($L_{\mathrm{bol}}/c$)  & 1.2&$<0.5$ \\
&& &  \\
\swift\ 2018 & $\dot M$ &$0.41\errUD{0.41}{0.33}$&$0.33\errUD{0.27}{0.20}$\\
& $\dot E$&$0.011\errUD{0.011}{0.008}$ &$0.047\errUD{0.038}{0.029}$\\
 &  log\,$L_{\mathrm{KIN}}$&44.1&44.8\\
&$\dot p_{\mathrm {w}}$ ($L_{\mathrm{bol}}/c$)    &1.2&2.3  \\
&&&\\
\it {ALMA$^d$}    &  $\dot M_{\mathrm{out}}$($M_\odot$yr$^{-1}$)&54 &\\
\it {ALMA}    &  log\,$E_{\mathrm{CO}}$&42 &\\
\it {ALMA}    &  $\dot p_{\mathrm {CO}}$($L_{\mathrm{bol}}/c$)& 0.8 &\\

    \hline
\end{tabular}
\end{table*}

In Table 3 we report  the resulting values for all the phases of the X-ray wind;  there we list  the outflow rate ($\dot M$) and  the kinetic output ($\dot E$) of the wind normalised  to the Eddington rates. The errors are calculated by propagating the 90\% errors on the column density and velocity of the disk wind  (see Table~ 2). It is clear that the  wind seen during the \suzaku\ observation,  with a higher $\dot M$  and $\dot E$, is  much stronger than during the \xmm\ slice\,A. On the other hand, the wind seen in the \chandra\  and \swift\ spectra, because of the large errors,   could be consistent with either of the  two previous epochs. 
While  all the derived  $\dot M$ are a substantial fraction of  the Eddington rate, the $\dot E$ are typically constrained in \sorg\ to within 1-3\% of Eddington, in line with the typical estimates for powerful winds in other AGN (e.g. \citealt{Gofford2015}). 
 
In order to compare the disk wind kinematics  with the ones reported for the possible kpc-scale molecular outflow (\citealt{Sirressi2019}), we  considered the values obtained for  the persistent component (zone 1) and converted the $\dot E$ to   the absolute values of the  kinetic output by  adopting the     estimated black hole mass of \sorg\ of $M_{\mathrm{BH}}\sim  10^{8}$\,M$_\odot$ (see B18 and  \citealt{Sirressi2019}). The kinetic power of this zone ranges from  $L_{\rm {KIN}}\sim  10^{44}$ erg s$^{-1}$, as measured during the  first part of the \xmm\ observations, to $\sim 3\times 10^{44}$\, erg  s$^{-1}$ for the \suzaku\  observation (see Table~3).
Thus on average the   derived kinetic output of this zone  is of the order of $L_{\rm {KIN}}\sim 1.7\times 10^{44}$\,\lum, which is  $\sim 5 \% L_{\mathrm{bol}}$, where  the  bolometric  luminosity   of \sorg\ is $L_{\mathrm{bol}}\sim 3\times 10^{45}$\,\lum  (B18). Regardless of the $\nhsym$ variability, all the estimates of the  kinetic energy of the X-ray wind are at least two orders of magnitude above  the kinetic  energy  carried  by  the   molecular gas phase (with $\dot E_{\rm CO}/\dot E_{\rm X}$ ranging from $\sim 3\times 10^{-3}$ to $\sim  10^{-2}$; \citealt{Sirressi2019}).  Note that, despite the large errors on  the $\nhsym$ and  the uncertainties on the black hole mass, we cannot reconcile the  different measurements of the X-ray wind energetics versus the large scale CO wind. Even if we consider the lowest possible measurement of $\dot E \sim 0.003$ (e.g the 90\% lower value for the \swift\ observation, see Table 3) we derive $L_{\rm {KIN}}\sim 3.8\times 10^{43}$\,\lum. Therefore  we would need a $\sim 40$ times smaller black hole mass (e. g. $M_{\rm{BH}}\sim 5 \times 10^6 M_\odot$) to have  $L_{\rm {KIN}}\sim  10^{42}$\,\lum in order to be compatible with the CO wind kinetic power.   This is implausible, because such a low black hole mass would make \sorg\ a super Eddington source, given its bolometric luminosity of  $L_{\mathrm{bol}}\sim 3\times 10^{45}$\,\lum. \\

  The momentum rate  of the X-ray wind  is $\dot p_{\rm {out}} =   \dot M_{\rm {out}}\, v_{\rm{out}}$;  here $\dot  p_{\rm {out}}$ was calculated again by adopting  $M_{\mathrm{BH}}\sim  10^{8}$\,M$_\odot$. The momentum rates  of the disk  wind  listed  in Table~3 are  all  normalised to  the momentum rate of the radiation $\dot p_{\mathrm {rad}}=L_{\mathrm{bol}}/c$.   Hence the X-ray wind momentum rate ranges from  $\dot p/ \dot p_{\mathrm {rad}}\sim 0.9$ (slice\,A) to $\sim 2.7$ (\suzaku\ observation).  Although these estimates are clearly affected  by large uncertainties,  the momentum rates  derived in  each of the observations is  of the same order of the radiation momentum rate (see Table 3), suggesting that this stream of the wind could be  radiation driven  with only a moderate  force multiplier.

We note that  the momentum rate  of the molecular outflow is consistent with that of  the X-ray wind suggesting that the  outflow is more likely to  be momentum rather than energy conserving on large scales. A  similar scenario also applies to other powerful X-ray winds such as PDS\,456 (\citealt{Bischetti2019}), IZw 1 (\citealt{ReevesBraito2019}), IRASF\,11119 + 3257 (\citealt{Veilleux2017,Nardini_Zubovas18}) and the  high redshift QSO, APM\,08279 + 5255 (\citealt{Feruglio2017}).  We note that recent hydrodynamical simulations of small scale winds in galaxies  predict larger scale energy conserving winds, because the Compton cooling times are expected to be short compared to the dynamical times of the inner winds (\citealt{Costa2020}),  meaning the solely momentum-driven wind can hardly happen.  It is possible that a  more energetic large scale outflow is present in \sorg,   but  it is   ionized rather than molecular gas. Interestingly, in  the optical spectrum  from  the 6dF Galaxy Survey (6dFGS, \citealt{Jones2009}), there is evidence for  possible blue wings in the  profiles of the [O III]$\mathrm{\lambda 4959, \lambda5007}$~\AA{}  emission lines  blue-shifted by $v_{\rm{ out}}\sim 500$\,km s$^{-1}$;  unfortunately   from the currently available   spectrum we cannot derive an estimate on   the  morphology and energetics of this possible ionised large scale outflow.  Future spatially resolved optical spectroscopy  using the VLT  Multi-Unit Spectroscopic Explorer  (MUSE) could  reveal the structure and kinematics of this ionised  component of the large scale outflow  and provide an estimate its energetics, which could  then be  compared  with the X-ray and large scale molecular winds.  \\

 We would like to stress the following consideration; had we not witnessed that  the slice\,B spectrum is due to  a   rare occultation event, as if for example  this was the only observation of the wind, we would have inferred an  incomplete  picture of  the disk wind. First of all, we note that  with the finer grid  we now measure a  less extreme wind parameters of  $\dot M\sim 0.31$ and $\dot E\sim 0.026$, which are similar  to the values measured during the \suzaku\ observation.  However, if the  wind is at the escape radius, which can be derived from  $v_{\rm{out}}\sim 0.13\,c$ ($R\sim 10^{15}$\,cm), its ionisation should be  higher, of the order of   log($\,\xi /{\rm erg\,cm \,s^{-1})}\sim 5$. This can be derived using the relation $R^2=L_{\rm{ion}}/n_{\mathrm {e}}\xi$ \footnote{where $L_{\rm{ion}}$ is the ionising luminosity in the 1-1000 Rydberg range ($L_{\rm{ion}}\sim 10 ^{45}$\, \lum\ as derived in B18) and $n_{\mathrm {e}}$ is density of the gas} assuming that the size of the absorber is $\Delta R/R=0.1$. 
  However, if we adopt the more realistic location of the wind of $R\sim 10^{16}$\,cm, as  derived by B18  from the absorption variability, we  infer a much higher mass outflow rate ($\dot M_{\rm {out}} \sim 4\,M_\odot /{\rm yr}$). The   kinetic output would    be     of the order of  10\% of  $L_{\rm{Edd}}$.  These more extreme  values are likely implausible and the transient nature of the slice B absorption suggests  that this is due to a passing wind clump of higher than usual density (and consequently lower ionisation). Such clumps likely have a lower filling factor which is confirmed by the fact that this state is seen only once among all the 11 observations (see Table 1) that we have for \sorg. We note that no spectral variability was seen amongst all the \swift\ observations, whereas   an obscuration event like the one seen in slice\,B, would have been easily detected. In particular, although  we could not characterise the wind in each of the short \swift\ observations, we would have been able to detect the strong curvature imprinted by this denser patch  in the 3-6 keV band. This stresses even more the importance of multi epoch observations to understand the nature of the X-ray winds.
  
\subsection{Faster outflow components}
Regarding the  two faster zones of the X-ray wind, which are outflowing with  $v_{\mathrm {out2}}\sim- 0.2\, c$ and  with   $v_{\mathrm {out3}}\sim -0.35\, c$, a possible scenario is  that they are   innermost streamlines of the wind closer to the black hole. In the discussion below, we concentrate  on the $ 0.2\,c$ outflow, which has been independently  confirmed in two of the observations.

For the  component    outflowing with   $v_{\mathrm {out2}}\sim -0.2\, c$, from its velocity we can infer that  this zone is most likely launched at a distance of $R_{\rm{w2}}\sim 7.5 \times 10^{14}$\, cm (or $\sim 50\, R_\mathrm {g}$);   this implies that the density  has to be of the order of $n_{\rm {e}}\sim  10^{10}$\, cm$^{-3}$  so  that  the iron is not fully ionized and we are able to detect it. This in turn implies a rather small size scale for the streamline or clumps of the wind  of $\Delta R=\nhsym/n_{\mathrm{e}}\sim 10^{14}$\,cm, which   is comparable to the X-ray source size ($\sim 10\, R_{\mathrm{g}}$).
Thanks to the multi epochs observations,  it  is now  clear that zone 2 is  sporadic; indeed,  we can place a rather stringent upper  limit on the $\nhsym$ of this zone to be less than $1.5\times 10^{23}$\,\nh\ during the \chandra\ observation (see Fig.~\ref{fig:bestfitCH_SW}). 

 The appearance of this  zone seems to be irregular  and uncorrelated to  the X-ray luminosity. A possible scenario is  that this faster zone is always present but, being launched  from closer in,  is generally more ionized and we are able to detect it only when our line of sight intercepts, like in slice\,B,  a denser  and/or lower ionisation clumps.  Alternatively it is possible that  this inner streamline  corresponds to different   ejecta events. 
In terms of its energetics, while its mass outflow rate is smaller or comparable to the mass carried by the slower phase of the wind,  the kinetic output  is higher given the higher velocity and could be of the order of $\sim 6 \times 10^{44}$\, \lum, which  correspond to $\sim 20$\%  of $L_{\mathrm{bol}}$.  Taken at face value this would imply that this  phase could be even  more important in terms of the feedback on the host galaxy. Nonetheless, as this phase could be intermittent,  the derived kinetic output, although apparently high,  may not be important in terms of its feedback as  the fast component could be a short lived phase.  

 We have very recently performed  a new series of four simultaneous \xmm\ \&  \nustar\ observations of \sorg, covering a time period of a month, that will allow us to probe the short term variability of the wind and if and how it reacts to changes of the X-ray source luminosity. These observations will crucially allow us also to probe  the connection between the two faster zones of the wind  (Braito et al. in prep.).
 
  \section{acknowledgements}
We thank the referee for his/her useful comments that improved the  paper.  This research has made use of data obtained from  \suzaku, a collaborative mission between the space agencies of Japan (JAXA) and the USA (NASA). Based on observations obtained with \xmm, an ESA science mission with instruments and contributions directly funded by ESA Member States and the USA (NASA). This work is based on observations obtained with the Neil Gehrels \swift\ Observatory. This work has been partially supported by the ASI-INAF program I/004/11/4. This work made use of data from the \nustar\ mission, a project led by the California Institute of Technology, managed by the Jet Propulsion Laboratory, and funded by NASA. This research has made use of the NuSTAR Data Analysis Software (NuSTARDAS) jointly developed by the ASI Science Data Center and the California Institute of
Technology.  VB acknowledges  financial  support  through   the NASA  grant   80NSSC20K0793 and the Chandra grant GO7-18091X.  VB, RDC, PS and RS  acknowledge financial contribution from the agreements ASI-INAF n.2017-14-H.0.  GAM is supported by ESA research fellowships.

\section{Data Availability}
The  data used for this research  can be  accessed at  https://heasarc.gsfc.nasa.gov/docs/archive.html.

\end{document}